\documentclass[superscriptaddress,twocolumn]{revtex4}
\usepackage{color,epsfig,amsmath,graphicx,amssymb}
\usepackage[colorlinks=fals,linkcolor=blue]{hyperref}
\begin{document}

\title{Boundary behaviours of open vesicles in axisymmetric case}

\author{Xiaohua Zhou}
\email{zhouxiaohua@xijing.edu.cn; xhzhou08@gmail.com}
 \affiliation{Department of Physics, Xijing University, Xi'an 710123, China}
 \affiliation{Engineering Technology Research Center for Controllable Neutron Source and Its Application, Xijing University, Xi'an, 710123, China}

\begin{abstract}
A continuous transformation from a closed vesicle to an open vesicle needs that the area of open hole enlarge from zero. Since the shape equation and boundary conditions of lipid open vesicles with free edges have been obtained, we want know whether
this process can be archived with valid parameters. By study the boundary conditions in the axisymmetric case, the analytic expression of the boundary edges is obtained generally. It reveals that the radius and line tension of boundary edges are confined strongly by bending moduli. In some cases, there is the minimal nonzero boundary radius and the line tension needs to surmount the maxim following the increase of boundary radius. Without the spontaneous curvature, the line tension will trend to infinite when the boundary radius shrinks to zero. The continuous opening process from a closed vesicle to an open vesicle needs that the spontaneous curvature is nonzero and the ratio between the bending moduli of Gauss curvature and mean curvature satisfies $k\leq-4$. Besides the line tension, a closed vesicle can be opened through changing the spontaneous curvature.

\end{abstract}
 \maketitle
\section{introduction}

Bilayer membranes are composed by two layers of phospholipid molecules which are the main components of cell membranes. In water environment, bilayer membranes can form closed vesicles \cite{Szoka1980, Evans1990PRL, Hotani1984JMB, Jung2002PNAS}. When increasing the protein concentration of the environment, the closed vesicles can change to open vesicles. Inverse deformation will occur when decreasing the protein concentration \cite{Saitoh1998PNAS}. Studying the equilibrium shapes and phase evolving paths of vesicles will give some insights of the relationship between the physical properties of cells and their biological functions.

When ignore the cytoskeleton, cells can be taken as bilayer membranes. Besides the artificial bilayer membranes which have relatively simple components, the red blood cell which has no nucleus are goods examples to test theoretical models. Early theoretical studies are difficult to explain experiments, until Helfrich pointed that membranes are in the liquid crystal state. In this model the membranes are taken as two-dimensional (2D) surfaces and their equilibrium shapes are determined by minimization of the free energy \cite{Helfrich1973}.
The energy density is a quadratic form of the two main cultures of membranes. Also, this model is named the spontaneous curvature (SC) model because it contains a spontaneous curvature which probably derives from the nonsymmetric factors of the inner and outer of the vesicles, such as the electric field intensity on membrane \cite{Meyer1968PRL, David1998TL}. Ou-yang and
Helfrich obtained the general shape equation of closed
membranes for the SC model \cite{OuYang1987PRL, OuYang1989PRA}. Several special analytical solutions have been found, such as the Clifford torus \cite{OuYang1990PRA}, discounts \cite{Naito1993PRE}, the
beyond-Delaunay surface \cite{Naito1995PRL} and the general solution for cylindrical case \cite{ZhangSG1996PRE, ZhangSG1997APS}. Particularly, the discounts solution is close to the red blood cell \cite{Naito1993PRE}. More details of these solutions are shown in Ref.\cite{OuYangBook}.
Besides, numeric solutions in the axisymmetric case have been studied extensively and the phase diagram was obtained \cite{Seifert1991Phase}. Many non-axisymmetric solutions and the corresponding stable phases ware obtained by using the finite element method \cite{YanJ1998PRE,Michalet2007PRE,ZhouXH2008IJMPB}. Based on the SC model, other models have been developed, such as bilayer-coupled (BC) model \cite{ Evans1980BJ} and the area-difference-elasticity (ADE) model \cite{Miao1994PRE}. In these two models, the thickness and area change of membranes in bending processes are considered. But it is very difficult to find analytic solutions for the equilibrium shape equations and some numeric solutions were studied \cite{Seifert1991Phase, Miao1994PRE, Ziherl2007PNAS}.

Talin is a ubiquitous mechanosensitive protein and plays an important role in cell actin and adhesion \cite{Kumar2016JCB}. Experiments have found that closed vesicles will change to open vesicles when increasing the talin concentration \cite{Saitoh1998PNAS}. In this process, talin assembles on the boundary edges and results in the deformation. Based on the Helfrich's free energy model, many researchers discussed the equilibrium shapes of open vesicles with free edges \cite{YinYJ2005CLSB, TuZC2003PRE, TuZC2010JCP, Biria2013AAM}. It has found that the shape equation for open vesicles are similar to closed vesicles with zero osmotic pressure. Besides, the edges should satisfy the boundary conditions \cite{YinYJ2005CLSB, TuZC2010JCP} which nearly eliminates the possibility to find a suitable boundary line on the known solutions of the shape equation of closed vesicles in the axisymmetric case \cite{TuZC2010JCP}. Up to know, only an integral case for the shape equation and boundary conditions was found when the line tension of the free edges can be negative  \cite{ZhangYH2018CPB,ZhouXH2018IJNLM}. Finding new analytic solutions for open vesicles is still a challenge \cite{TuZC2013CPB}.

The phenomena in Ref.~\cite{Saitoh1998PNAS} indicate that a closed vesicle can be opened continuously because that the area of open hole enlarge from zero. Since the shape equation and boundary conditions of open vesicles with free edges have been obtained, we want know whether this process can be achieved with valid parameters. For instance, at the beginning of the opening process of a vesicle with an infinite small hole, whether the line tension of the free edge is finite? If it is not, we think the opening process is discontinuous or cannot be achieved physically. In other words, we want to know wether it needs infinite forces to open a closed cell. Cells have many ion channels which are controlled by the membrane potential. If we take these channels as small open holes, whether the opening and closing actions can be achieved in the open vesicle model?
Moreover, in the axisymmetric case the boundary conditions are reduced to ordinary deferential equations. In previous works, researcher tried to find solutions of the shape equation of open vesicle with these boundary conditions. Although this method provides some special shapes, revealing the insights of the general behaviours of open vesicles is still difficult. However, if these boundary equations can be solved generally, we can obtain the general characteristics of the boundary edges and avoid to solve the shape equations.

In this work we will study the above questions by investigating the deferential equations of boundary conditions in the axisymmetric case. Supposing the talin concentration is positively related to the line tension of the free edges, the general solution is obtained and it reveals some of the boundary behaviours of open vesicles which are in good agrement with experiments. In Sect. II, the boundary equations are simplified and the general solution is derived analytically.
In Sect. III, general boundary behaviours of open vesicles are discussed and the results are compered with experiments. Finally, these results are recapped in a short discussion in Sect. IV.

\section{Analytic solution for axisymmetrical boundary conditions}
The Helfrich free energy for a vesicle is \cite{Helfrich1973}
\begin{eqnarray}\label{Energy}
E=\frac{k_{c}}{2}\int\int(2H-C_{0})^{2} dA+k_b\int\int\Lambda dA,
\end{eqnarray}
where $k_{c}$ and $k_{b}$ are the bending
moduli, $H$ and $\Lambda$ are the mean curvature and Gauss curvature, respectively. $C_0$ is the spontaneous curvature and $dA$ is the area element. For an open vesicle with a free edge, the total energy is
\begin{eqnarray}
E_t=E+\lambda\int\int dA +\gamma L,
\end{eqnarray}
where $\lambda$ is the area tension and $\gamma$ is the line tension of the boundary edge with the length $L$.

For an axisymmetric surface, let the generating line be around the $z$-axis and $\rho$ be the turning radius,
\begin{equation}\label{xyz}
x=\rho \cos\phi, y=\rho \sin\phi, z=\int\tan\psi(\rho)d\rho,
\end{equation}
where $\phi$ is the azimuthal angle and $\psi$ is the tangent angle of the profile curve.
The surface can be expressed as
\begin{equation}\label{r}
 \textbf{r}=\{\rho \cos\phi,\rho \sin\phi, z \} .
\end{equation}
Defining $()'=\frac{d()}{d\rho}$, we have
\begin{eqnarray}
  \label{K1}&&K_{1}=\frac{\sin\psi}{\rho},~K_{2}=(\sin\psi)^\prime,\\
 \label{H1}&&H=(K_1+K_2)/2,~~\Lambda=K_1K_2,
\end{eqnarray}
where $K_1$ and $K_2$ are two main curvatures. Hu et. al obtained the general shape
equation, which is a third-order ordinary differential equation \cite{HuJG1993PRE}. Zheng et. al provided the first integral of the shape equation as follow \cite{ZhengWM1993PRE}.
\begin{eqnarray}\label{Eq1}
\bar{\lambda}=(2H-C_0)\psi^\prime\cos \psi+2H^\prime\frac{\cos^2\psi}{\sin\psi}-\frac{1}{2}(2H-C_0)^2,
\end{eqnarray}
where $\eta$ is a constant of integration, $\tilde{\lambda}=\lambda/k_c$ and $\tilde{p}=p/k_c$. The above is a second-order ordinary differential equation with the variable $\rho$. Although it is very difficult to solve generally, several analytic solutions have been found \cite{OuYangBook}.

 Besides, the edges should satisfy the following
boundary conditions \cite{Capovilla2002PRE,TuZC2003PRE}
\begin{eqnarray}
\label{condition1-1}k=-\frac{2H_c-C_0}{\sin\psi_c}\rho_c,\\
\label{condition1-2}\frac{1}{2}(2H_c-C_0)^2+k\Lambda_c+\bar{\lambda}+\sigma\bar{\gamma}\frac{\cos\psi_c}{\rho_c}=0,
\end{eqnarray}
where $k=k_b/k_c$ and the subscript $()_c$ means the corresponding value on the boundary line and we name $\psi_c$ the opening angle. Substituting Eqs.~\ref{H1}-\ref{condition1-1} into Eq.~\ref{condition1-2} and choosing $\sigma=1$, we get
\begin{eqnarray}
\label{gamma-1}\bar{\gamma} = -2\rho_c H_c^\prime\cot\psi_c.
\end{eqnarray}
Making use of Eq.~\ref{H1}, Eq.~\ref{condition1-1} is changed to
\begin{eqnarray}
\label{Beq1}(\sin\psi_c)^\prime+\frac{k+1}{\rho_c}\sin\psi_c-C_0=0.
\end{eqnarray}
This equation has the following general solution
\begin{eqnarray}
\label{psi-1}\psi_c=\arcsin\bigg(C\rho_c^{-k-1}+\frac{C_0}{k+2}\rho_c\bigg).
\end{eqnarray}
where $C$ is a constant of integration.

Defining the scale factor $\rho_0=|C|^{\frac{1}{k+1}}$, the reduced boundary radius $\tilde{\rho}_c=\rho_c/\rho_0$ and the reduced spontaneous curvature $c_0=C_0\rho_0$, we have
\begin{eqnarray}
\label{psi-2}\psi_c=\psi_{c1}\equiv\arcsin\bigg[\delta\tilde{\rho}_c^{-k-1}+\frac{c_0}{k+2}\tilde{\rho}_c\bigg].
\end{eqnarray}
where $\delta=1$ for $C>0$ and $\delta=-1$ for $C<0$. Here we should not that the function $\arcsin$ confines $-\frac{\pi}{2}\leq\psi_c\leq\frac{\pi}{2}$. Actually, besides Eq.~{\ref{psi-2}}, the opening angle can be
\begin{eqnarray}
\label{psi-3}\psi_c=\psi_{c2}\equiv\pi-\arcsin\bigg[\delta\tilde{\rho}_c^{-k-1}+\frac{c_0}{k+2}\tilde{\rho}_c\bigg].
\end{eqnarray}

Furthermore, we define the following reduced parameters: $\tilde{K}_{1c}=\rho_0 K_{1c}$, $\tilde{K}_{2c}=\rho_0 K_{2c}$, $\tilde{H}_c=H_c\rho_0$, $\tilde{\Lambda}_c=\Lambda_c\rho_0^2$, $\tilde{\gamma}=\bar{\gamma}\rho_0$ and $\tilde{\lambda}=\bar{\lambda} \rho_0^2$. Substituting Eq.~\ref{psi-2} and Eq.~\ref{psi-3} into Eqs.~\ref{K1}-\ref{Eq1} and \ref{gamma-1}, we obtain
\begin{eqnarray}
\label{K11}&&\tilde{K}_{1c}=\frac{c_0}{k+2}+\delta\tilde{\rho}_c^{-2-k},\\
\label{K22}&&\tilde{K}_{2c}=\frac{c_0}{k+2}-\delta(1+k)\tilde{\rho}_c^{-2-k},\\
\label{H}&&\tilde{H}_c=\frac{c_0}{k+2}-\frac{\delta}{2} k \tilde{\rho}_c^{-2-k},\\
\label{A}&&\tilde{\Lambda}_c=\frac{c_0^2}{(2+k)^2}-(1+k)\tilde{\rho}_c^{-2(2+k)}-\frac{\delta c_0 k}{2+k}\tilde{\rho}_c^{-2-k},\\
\label{gamma-2}&&\tilde{\gamma}=\epsilon\frac{\delta k(k+2)^2\sqrt{1-[c_0\tilde{\rho}_c/(k+2)+\delta\tilde{\rho}_c^{-k-1}]^2}}{\delta(k+2)\tilde{\rho}_c+c_0\tilde{\rho}_c^{3+k}},\\
\label{Lambda-1}&&\tilde{\lambda}=\frac{\delta k(2+k)^2\tilde{\rho}_c^{-2}}{\delta(2+k)+c_0\tilde{\rho}_c^{2+k}}-\frac{k\big[\delta(2+k)+c_0\tilde{\rho}_c^{2+k}\big]^2}{2(2+k)\tilde{\rho}_c^{2(2+k)}}.~~~~
\end{eqnarray}
Where $\epsilon=-1$ for $\psi_c=\psi_{c1}$ and $\epsilon=1$ for $\psi_c=\psi_{c2}$. Particularly, Eq.~\ref{K11} and Eq.~\ref{K22} yield
\begin{eqnarray}
\label{K1K2}\tilde{K}_{1c}(1+k)+\tilde{K}_{2c}=c_0.
\end{eqnarray}
This condition gives a general structure characteristic of the boundary edges.

The shape of open vesicles can be changed by adjusting the concentration of talin \cite{Saitoh1998PNAS}. A reasonable interpretation is that talin assembled boundary edges changes the line tension $\gamma$. But we do not know whether a closed vesicle can be opened by choosing suitable $\gamma$. In other words, we need $\gamma$ is finite at the beginning of the open process. For an open vesicle, if $\gamma\rightarrow\pm\infty$ or $\tilde{\lambda}\rightarrow\pm\infty$  when the area of open hole shrinks to zero, the deformation between the closed one and open one cannot be archived physically.

In the axisymmetrical case, on the north pole of a closed vesicle, there are $\psi=\pi$ and $\rho=0$.
Supposing that the opening process begins at this pole and the closed vesicle changes to a open vesicle continuously, the boundary edge satisfies
\begin{eqnarray}
\label{condition0}\tilde{\gamma}~\text{and}~\tilde{\lambda}~\text{are finite},~\psi_c\rightarrow\pi,~\text{when}~\tilde{\rho}_c\rightarrow0.
\end{eqnarray}
Moreover, Eq.~\ref{Eq1} needs that $h$ and $h'$ are also finite when boundary radius $\tilde{\rho}\rightarrow0$.
So, in the following text, we discuss whether the above conditions can be archived in Eqs.~\ref{gamma-2} and \ref{Lambda-1} by choosing suitable parameters

\section{Shape transformation from a closed vesicle to an open vesicle}
\subsection{Without the spontaneous curvature}
\begin{figure*}\centering
\includegraphics[height=12cm]{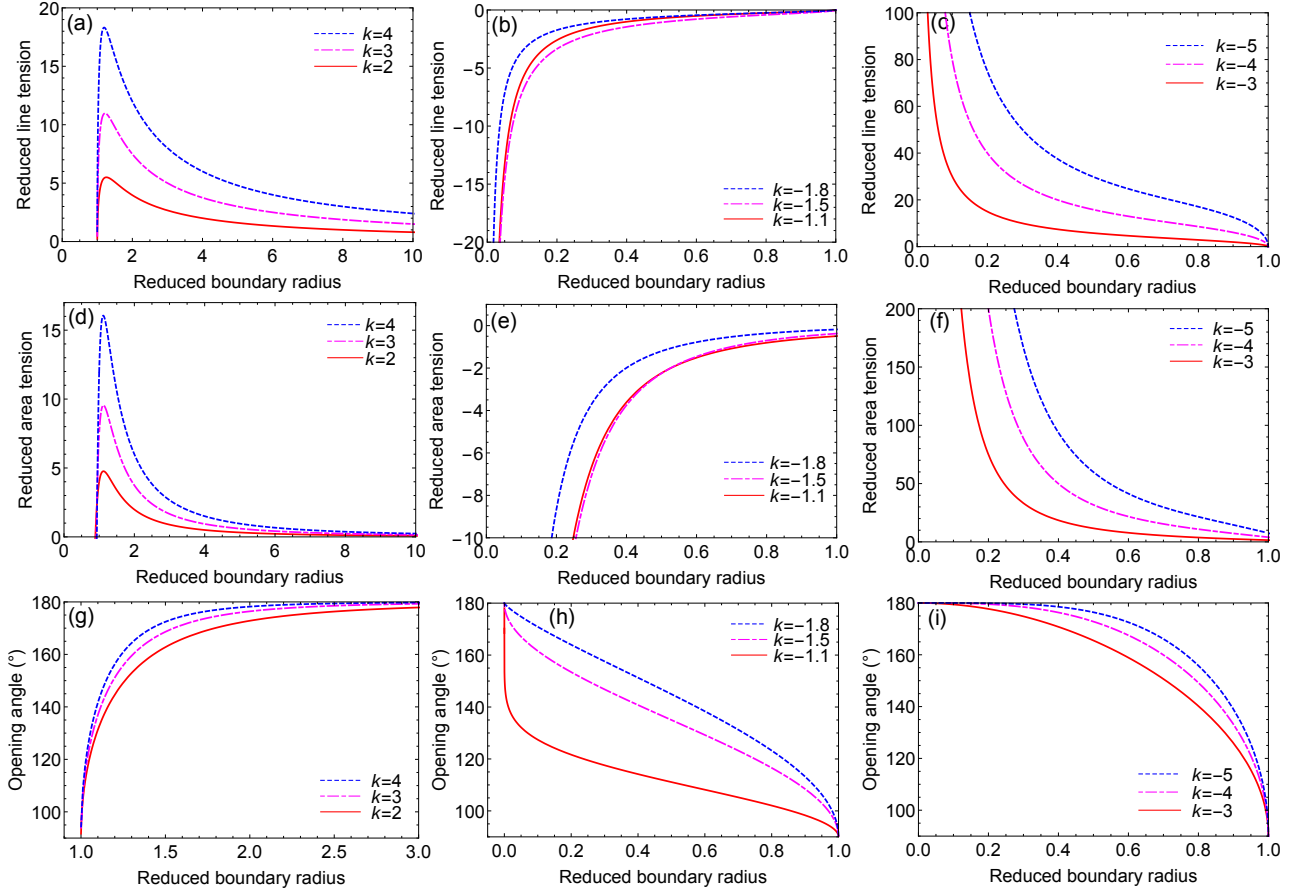}
\caption{ Relationships between the reduced boundary radius $\tilde{\rho}_c$, the reduced line tension $\tilde{\gamma}$ and the opening angle $\psi_c$. }
\label{fig1}
\end{figure*}

First, let us consider the state $c_0=0$. Then,  Eqs.~\ref{psi-1}, \ref{gamma-2} and \ref{Lambda-1} are reduced to
\begin{eqnarray}
\label{psi-c0}&&\psi_c=\arcsin(\delta\tilde{\rho}_c^{-k-1})~ \text{or}~\psi_c=\pi-\arcsin(\delta\tilde{\rho}_c^{-k-1}),~~~~~\\
\label{gamma-c0}&&\tilde{\gamma}=\epsilon k(k+2)\tilde{\rho}_c^{-1} \sqrt{1-\tilde{\rho}_c^{-2k-2}},\\
\label{Lambda-c0}&&\tilde{\lambda}=-\frac{1}{2}k(k+2)\big[\tilde{\rho}_c^{-2} -2 \tilde{\rho}_c^{-2(k+2)}\big].
\end{eqnarray}

In Eq.~\ref{gamma-c0} we need $1-\tilde{\rho}_c^{-2k-2}\geq0$ and it yields
\begin{eqnarray}
\label{c1}\tilde{\rho}_c\geq1,~~k>-1,\\
\label{c2}\tilde{\rho}_c\leq1,~~k<-1.
\end{eqnarray}
We can also obtain
\begin{eqnarray}
\label{s1}\tilde{\gamma}= 0,~\psi_c=\pm\pi/2,~\text{when}~\tilde{\rho}_c= 1~\text{and}~k>-1,\\
\label{s2}\tilde{\gamma}\rightarrow\pm\infty,~\psi_c=0~\text{or}~\pi,~\text{when}~\tilde{\rho}_c= 0~\text{and}~k<-1.~
\end{eqnarray}
Moreover, we can find that $\tilde{\lambda}\rightarrow\pm\infty$ when $\tilde{\rho}_c= 0$. Since $\tilde{\gamma}\rightarrow\pm\infty$ has no physical meaning, the above results indicates that a closed vesicle cannot changes to a open vesicle continuously when $c_0=0$. We show several examples of the relationships between the $\tilde{\rho}$, $\tilde{\gamma}$ and $\tilde{\rho}_c$ in Fig.~\ref{fig1}.

\subsection{With nonzero spontaneous curvature}
When $c_0\neq0$, the relationship between $\tilde{\gamma}$ and $\tilde{\rho}$ is intricate. Besides the behaviours in Fig.~\ref{fig1} , we can find the following phonomania ($\epsilon=1$)
\begin{eqnarray}
\label{s4}&&\tilde{\gamma}=0,~\psi_c=\pi,~\text{when}~\tilde{\rho}_c= 0~\text{and}~k<-3,\\
\label{s5}&&\tilde{\gamma}=3\delta/c_0,~\psi_c=\pi,~\text{when}~\tilde{\rho}_c= 0~\text{and}~k=-3.~~~
\end{eqnarray}
The above results satisfy the conditions in Eq.~\ref{condition0}. We also find that there always is $\gamma\rightarrow\pm\infty$, when $k>-3$. Moreover, considering that continuous deformation needs $\tilde{\lambda}$ is infinite. In Eq.~\ref{Lambda-1}, if $k>-4$, we find $\tilde{\lambda}\rightarrow\pm\infty$ when $\tilde{\rho}_c= 0$.
The valid states are
\begin{eqnarray}
\label{s6}&&\tilde{\lambda}=\frac{-kc_0^2}{2(2+k)},~\text{when}~\tilde{\rho}_c= 0~\text{and}~k<-4,\\
\label{s7}&&\tilde{\lambda}=-16\delta/c_0-c_0^2,~\text{when}~\tilde{\rho}_c= 0~\text{and}~k=-4.
\end{eqnarray}
Moreover, if $k\leq-4$, it is easy to get
\begin{eqnarray}
\label{h0} \tilde{H}_c=\frac{c_0}{2+k},~\tilde{\Lambda}_c=\frac{c_0^2}{(2+k)^2},~\text{when}~\tilde{\rho}_c= 0.
\end{eqnarray}

A closed spherical solution with the radius $R$ and zero pressure satisfies \cite{OuYang1987PRL}
\begin{eqnarray}
\label{sp} 2\tilde{\lambda} R-c_0(2-c_0R)=0.
\end{eqnarray}
Supposing that this sphere begins to open up with $\tilde{\rho}_c=0$, Eq.~\ref{h0} gives $R=1/\tilde{H}_c=(2+k)/c_0$. This result and $\tilde{\lambda}$ in Eq.~\ref{s6} satisfy Eq.~\ref{sp}. Therefor, the open process can begin with a sphere.

\begin{figure}\centering
\includegraphics[height=3cm]{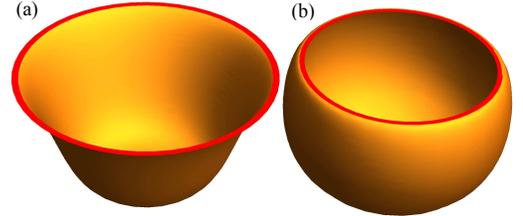}
\caption{ Two kinds of impossible shapes due to the infinite line tension on each boundary. There are $\psi_c=0$ for (a) and $\psi_c=\pi$ for (b).}
\label{fig2}
\end{figure}

In Eq.~\ref{psi-1}, we define
\begin{eqnarray}
\label{f} f=\delta\tilde{\rho}_c^{-k-1}+\frac{c_0}{k+2}\tilde{\rho}_c.
\end{eqnarray}
If $f=0$, we have $\psi_c=0$ or $\psi_c=\pi$. Also, it yields $\rho_c=0$ and $\rho_c=(\frac{k+2}{-\delta c_0})^{\frac{1}{k+2}}$. But if $\rho_c=(\frac{k+2}{-\delta c_0})^{\frac{1}{k+2}}$, we have $\tilde{\gamma}\rightarrow \pm\infty$. Therefor, this result gives two kinds of impossible shapes as shown in Fig.~\ref{fig2}. Moreover, $f=\pm1$ yields $\psi_c=\pm\pi/2$ and $\gamma_c=0$.

Based on the above analysis, we can conclude that the continuous deformation between closed vesicles and open vesicles can only occur when $c_0\neq0$ and $k\leq-4$.

\subsection{Analytic results Compered with experiments }

Experiments have revealed that a closed vesicle can be opened through increasing the tain concentration and this process seems to be continuous \cite{Saitoh1998PNAS}. Inversely, decreasing the taling concentration induces that a open shape changes to a closed one. In these deformations, talin assembles on the boundary edges and changes the mechanical state of vesicles. A reasonable assumption is that the tain concentration is positively related to the reduced line tension $\tilde{\gamma}$. Then, increasing $\tilde{\gamma}$ can open a closed vesicle and inverse process will occur when decreasing $\tilde{\gamma}$. Choosing suitable parameters, these process can be archived theoretically.

\begin{figure}\centering
\includegraphics[height=11cm]{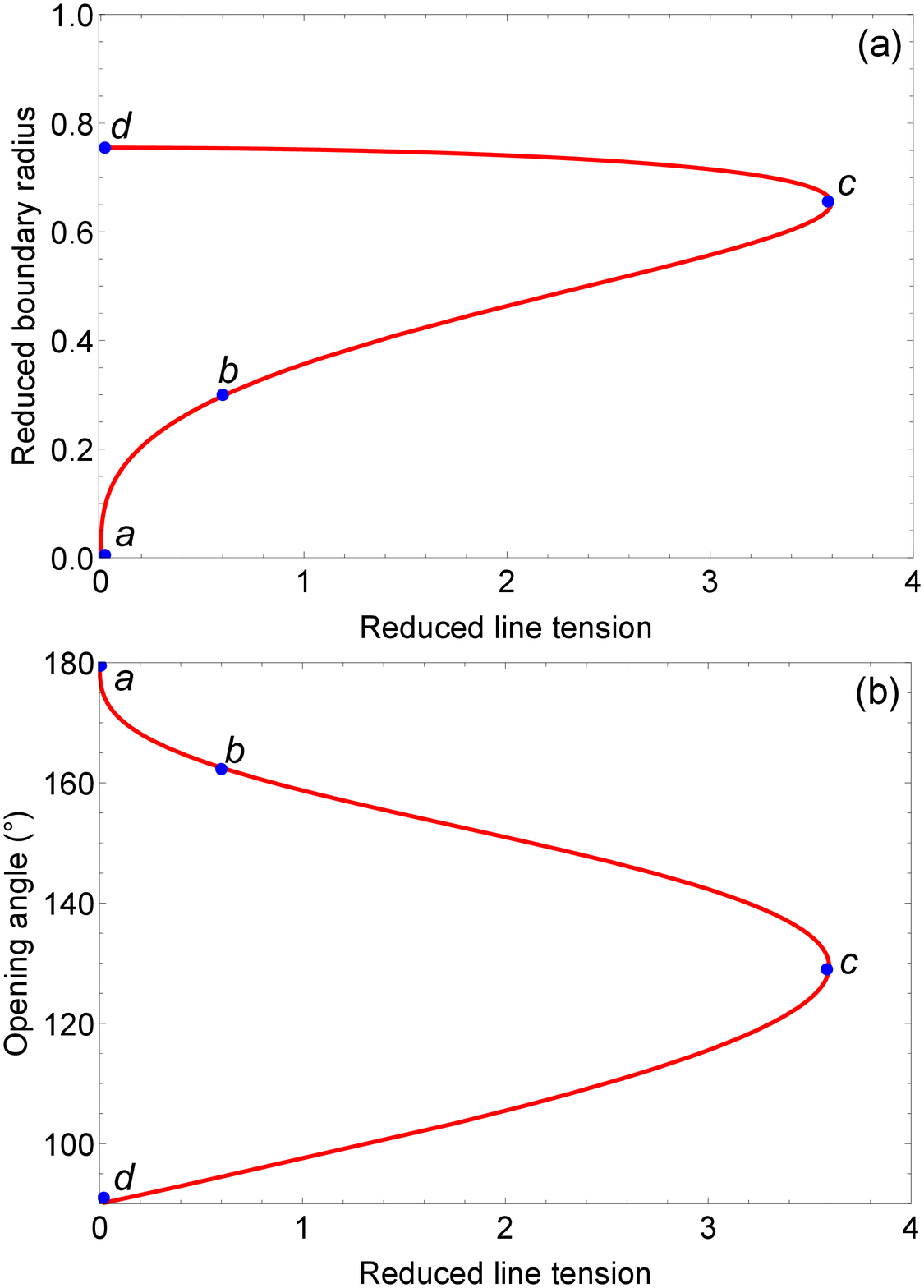}
\caption{ The relationships between the opening angle $\psi_c$ and reduced bounder radius $\tilde{\rho}_c$, and between reduce line tension $\tilde{\gamma}$ and $\tilde{\rho}_c$. Here we chose $k=-6$, $c_0=-4$, $\psi_c=\psi_{c2} $ and $\delta=1$. In each chart, there are $\tilde{\rho}_c=0$ and $\psi_c=180^\circ$ for point $a$,   $\tilde{\rho}_c=0.3$ and $\psi_c=162.3^\circ$ for point $b$, $\tilde{\rho}_c=0.3$ and $\psi_c=129^\circ$ for point $c$, $\tilde{\rho}_c=0$ and $\psi_c=90^\circ$ for point $d$.}
\label{fig3}
\end{figure}
Fig.~\ref{fig3} shows an example by choosing $k=-6$, $c_0=-4$, $\psi_c=\psi_{c2} $ and $\delta=1$.
In Fig.~\ref{fig3}(a), the open process begins at the point $a$. With the increase of the reduced line tension $\tilde{\gamma}$, the reduced bounder radius $\tilde{\rho}_c$ will increase until it reaches the point $c$. On which, the open hole reaches the biggest state. Accordingly, Fig.~\ref{fig3}(b) shows that the opening angle decrease from $\psi_c=180^\circ$ on point $a$ to $\psi_c=162.3^\circ$ on point $c$. But if we keep on increasing $\tilde{\gamma}$ on point $c$, there is no evolving path for the shape. Considering that our model is in the axisymmetric case, we think the vesicle will change to non-axisymmetric shapes in this case.

\begin{figure}\centering
\includegraphics[height=4cm]{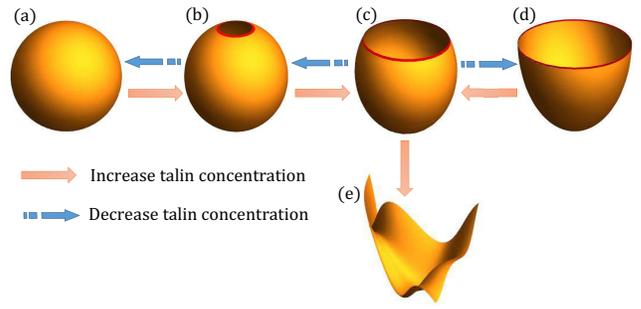}
\caption{ Deformation paths depend on the change of talin concentration. The (a)-(d) shapes are corresponding to the $a$-$d$ points in Fig.~\ref{fig3}. There is $\psi_c=\pi$ for (d) and shape (e) is a non-axisymmetric shape.}
\label{fig4}
\end{figure}
Moreover, if we decrease $\tilde{\gamma}$ on point $c$, there are two deformation paths. One changes back to the initial closed shape on point $a$. The other changes to point $d$ with a nonzero $\tilde{\rho}_c$. Fig.~\ref{fig4} shows the evolving paths in the 3D case. The (a)-(d) shapes are corresponding to the $a$-$d$ points in Fig.~\ref{fig3}. The process from (a) to (e) is in good agreement with experiments \cite{Saitoh1998PNAS}. Besides, we find a new deformation path between shapes (c) and (d), which needs to be proved in future experiments.

\subsection{Opening process induced by spontaneous curvature}
\begin{figure}\centering
\includegraphics[height=11cm]{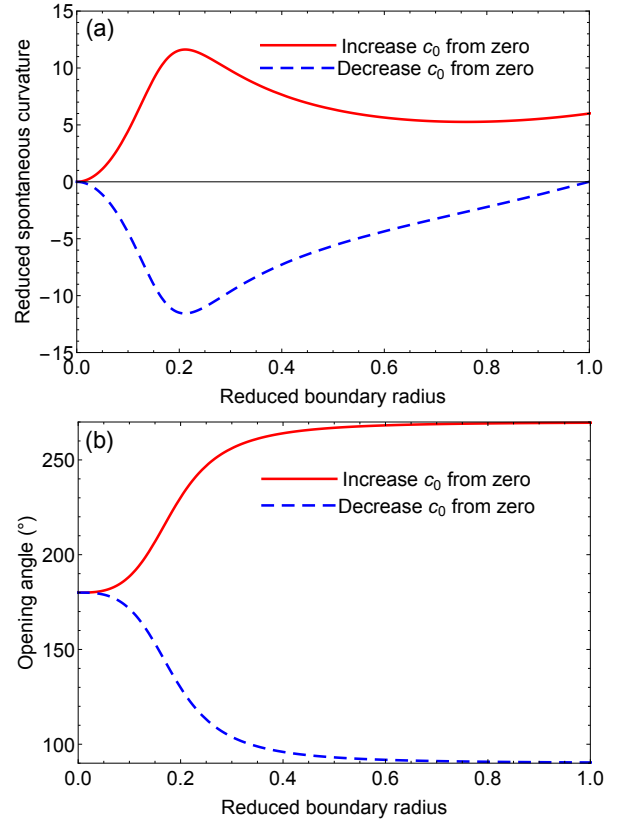}
\caption{ Relationships between the reduced bounder radius $\tilde{\rho}_c$ and spontaneous curvature $c_0$ and between $\tilde{\rho}_c$ and opening angle $\psi_c$. We fixed $\tilde{\gamma}=0.1$, $k=-5$ and $\delta=1$. }
\label{fig5}
\end{figure}
\begin{figure}\centering
\includegraphics[height=3.2cm]{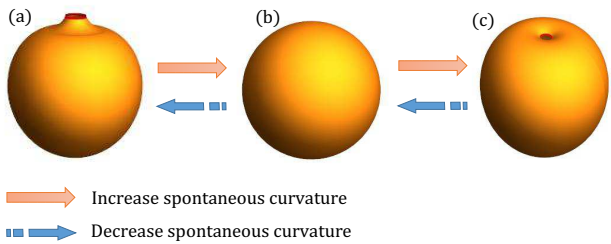}
\caption{ Opening and closing processes induced by the change of spontaneous curvature. The closed shape (b) with $c_0=0$ has an infinite small hole. Decreasing or increasing $c_0$ will induce which changes to shape (a) with a hole opened outward or shape (c) with a hole opened inward.}
\label{fig6}
\end{figure}

In the up sections we take $c_0$ and $k$ as constants and show that one can open a vesicle through increasing the line tension. But we don't known wether it is possible to open a closed vesicle by changing other parameters, such as $c_0$. Cells have many channels which are controlled by the membrane potential. It has reported that the $c_0$ is related to the electric field on membrane \cite{Meyer1968PRL, David1998TL}. So, it implies that $c_0$ is related to the membrane potential. Therefor, taking these channels as small open holes, opening and closing actions probably can be achieved by adjusting $c_0$.

Supposing that a channel on a cell is closed, then the cell can be taken as a closed vesicle with an infinite small hole. If no talin and other proteins are added in the vesicle environment, we think the reduced line tension $\tilde{\gamma}$ is a constant. In this case, if the membrane potential is changed chemically, $c_0$ will change accordingly. If $\tilde{\gamma}$ in Eq.~\ref{gamma-2} is fixed and taking $c_0$ as variable, we obtain the relationships between $c_0$ and $\psi_c$ and between $c_0$ and $\tilde{\rho}_c$. Fig.~\ref{fig5} shows an example by choosing $\tilde{\gamma}=0.1$. From Fig.~\ref{fig5}(a) we can see that the closed vesicle with $c_0=0$ is opened up through changing $c_0$. Increasing or decreasing $c_0$ from zero seems has similar effect. But there is also deference. Following the increase of $\tilde{\rho}_c$, there are the maximal and minimal points for $c_0$. Fig.~\ref{fig5}(b) indicates that increasing $c_0$ will enlarge the opening angle, which means that the hole is opened outward. Inversely, decreasing $c_0$ will reduce the opening angle and make a hole opened inward.
This deference is shown in Fig.~\ref{fig6} in the 3D case. Before $c_0$ reaches the maximal or minimal value, changing back $c_0$ to zero will close the open hole. This result implies that the opening and closing actions of channels can be achieved by changing membrane potential, which is a well known biological phenomenon.

\section{Conclusions} \label{Conclusions}

Opening up a hole and closing a channel are typical cell behaviors and they play an important rule in vital activities. Using continuum mechanics to study these phenomena, especially the opening and closing actions of ion channels, is facing on challenge.

In this work, based on the Helfrich free energy model, we obtain the general solution of the boundary equations of open vesicles in the axisymmetric case. Some general behaviours of the the boundary edges are revealed. The deformation paths are investigated, which are in good agreement with experiments. The results also indicate that the radius and line tension of boundary edges are confined strongly by bending moduli. Without the spontaneous curvature, the line tension will trend to infinite when the boundary radius shrinks to zero. Which implies that a closed vesicle cannot be opened by using finite forces in this case, or that there is discontinuous transformation between a close and a open vesicles. With finite physical parameters, opening and closing processes need that the spontaneous curvature is nonzero and that the ratio between the bending moduli of Gauss curvature and mean curvature satisfies satisfies $k\leq-4$. Taking the ion channels as open holes, the opening and closing actions can be manipulated by changing $c_0$. Since $c_0$ is related to the membrana potential, it implies that these actions can be achieved by changing membrane potential, which is in accordance with electrophysiology experiments.

This work is funded by
the National Natural Science Foundation of China Grants
(No. 11304383), the Natural Science Basic Research Plan of Shaanxi Province of China (No. 2018JM1019), and the scientific research plan projects of education department of Shaanxi provincial government (No. 17JK1155).

\end{document}